\documentstyle[12pt]{article}

\begin{document}

\setcounter{page}{1}

\pagestyle{plain}

\begin{center}
\Large{\bf Black hole thermodynamics in Snyder phase space }\\
\small \vspace{1cm} {\bf Sara Saghafi\footnote{s.saghafi@stu.umz.ac.ir}}\quad and \quad  {\bf Kourosh
Nozari\footnote{knozari@umz.ac.ir(Corresponding Author)}}\\
\vspace{0.5cm} Department of Physics, Faculty of Basic Sciences,
University of Mazandaran,\\
P. O. Box 47416-95447, Babolsar, IRAN
\end{center}

\begin{abstract}
By defining a noncommutative symplectic structure, we study thermodynamics of Schwarzschild black hole in a Snyder noncommutative phase space for the first time.
Since natural cutoffs are the results of compactness of symplectic manifolds in phase space, the physics of black holes in such a space would be affected mainly by these cutoffs.
In this respect, this study provides a basis for more deeper understanding of the black hole thermodynamics in a pure mathematical viewpoint. 
\end{abstract}
\newpage

\section{Introduction}
General Relativity has enhanced our insight on the universe. However, when dealing with describing the physical phenomena at high energy regime or very short distances, the problems and defections of this theory will come into play, especially in the order of the Planck scales where quantum effects as well as gravitational effects are important. It is then natural to expect that these problems will be addressed in the context of ultimate quantum theory of gravity. In the absence of such a conclusive theory, quantum gravity candidates such as string theory and loop quantum gravity suggest the existence of a minimal length scale (and also a maximal energy scale) leading to a universal ultraviolet (UV) cutoff [1,2]. Existence of these natural cutoffs is essential for the regularization of quantum field theories [3]. Evidently, taking a minimal length scale into account naturally makes quantum field theories to be UV-regularized [4-8]. Also, it is shown that the existence of an IR cutoff is necessary for the quantization of fields in curved spacetimes [9]. Thus, modification of general relativity in the high and low energy regimes seems to be necessary in order to take into account these natural quantum gravity cutoffs. Indeed, noncommutative geometry is an appropriate framework to formulate theories which deal with UV and IR cutoffs [10] (see Refs. [6,11-13] for the UV/IR mixing effect). The others effective phenomenological models of quantum gravity such as the generalized uncertainty principle (which is inspired by string theory [7,8,14]) and polymer quantum mechanics [15,16] (which is investigated in the symmetric sector of loop quantum gravity) are in close relation with noncommutative spaces (see for instance [17,18]). At the classical level, all of these models, and any model which includes natural UV and IR cutoffs, can be realized from a deformed Hamiltonian system. Such systems are usually led to deformed noncanonical Poisson algebras with non-vanishing commutation relations between positions and momenta which signal the existence of UV and IR cutoffs respectively [19]. This is, however, a local criterion and we know that the Hamiltonian systems are described by the symplectic manifolds which are locally equivalent. In Ref. [20] the authors considered the Hamiltonian systems in their most fundamental forms by using symplectic manifolds and showed that what is the origin of natural cutoffs. In 1947 Snyder started to establish the noncommutative geometry which today is known as Snyder noncommutative spacetime and its main focus was on noncommutativity on symplectic manifolds [21]. Symplectic manifolds are based on phase space coordinates and a symplectic structure is defined on it. Many attempts have been done to write the classical dynamics and quantum mechanics of Snyder space [22-24].  Also Snyder dynamics in Schwarzschild spacetime has been studied in Ref. [25]. In this letter, thermodynamics of black holes in Snyder spacetime is studied in a symplectic manifold. Since natural cutoffs are the results of compactness of symplectic manifolds in Snyder spacetime, the physics of black holes in such a space would be affected mainly by these cutoffs. Firstly we have a short section which shows how one can consider a Hamiltonian model for a spherically symmetric spacetime. Next, formulation of physics of black holes is presented in a commutative symplectic manifold. Then we consider this symplectic manifold to be a noncommutative one such as the Snyder noncommutative phase space, and the physics and thermodynamics of black holes are discussed in this setup for the first time. By calculating the partition function in noncommutative Snyder symplectic manifold, we study thermodynamical properties of the final phase of micro black holes in noncommutative phase space.

\section{A Hamiltonian model for Black Hole}

In Ref. [26] a Hamiltonian quantum theory of spherically symmetric, asymptotically flat electrovacuum spacetime is considered. The physical phase space of such a spacetime is spanned by the mass $M$ and the charge parameter $Q$ of the Reissner-Nordstr\"{o}m black hole together with the corresponding canonical momenta as
\begin{eqnarray}
H = \left( {{N_ + } + {N_ - }} \right)m + \left( {{\phi _ + } - {\phi _ - }} \right)Q\,,
\end{eqnarray}
where $N_ \pm$ and $\phi _\pm$ are the lapse functions and the electric potentials. In this four-dimensional phase space, they performed a canonical transformation such that the resulting configuration variables describe the dynamical properties of Reissner-Nordstr\"{o}m black holes in a natural manner. The classical Hamiltonian written in terms of these variables and their conjugate momenta is replaced by the corresponding self-adjoint Hamiltonian operator from the point of view of a distant observer at rest
\begin{eqnarray}
\begin{array}{l}
H\left( {{P_m},m} \right) \to H\left( {{P_a},a} \right)\\ 
H\left( {{P_a},a} \right) = \frac{{P_a^2}}{{2a}} + \frac{a}{2} + \frac{{{Q^2}}}{{2a}}\,.
\end{array}
\end{eqnarray}
By setting $Q=0$, the Hamiltonian of the Schwarzschild black hole is obtained. Now by using the functions of the configuration variable $a$ and substituting ${P_a} \to -i\frac{d}{{da}}$, the symmetric Hamiltonian operator is obtained as
\begin{eqnarray}
\hat{H} :=  - \frac{1}{2}{a^{ - s}}\frac{d}{{da}}\left( {{a^{s - 1}}\frac{d}{{da}}} \right) + \frac{a}{2}\,.
\end{eqnarray}

The Hilbert space is $H:= {L^2}\left( {{\Re ^ + },{a^s}da} \right)$, where $s$ is the factor ordering parameter. With a particular choice $s=2$ and identifying ${R_ \pm } = M \pm \sqrt {{M^2} }$ we have
\begin{eqnarray}
\frac{1}{a}\frac{{{\partial ^2}\psi }}{{\partial {a^2}}} + \frac{1}{{{a^2}}}\frac{{\partial \psi }}{{\partial a}} = \left( {a - {R_ + }} \right)\left( {a - {R_ - }} \right)\psi\,.
\end{eqnarray}

So, the quantized mass of the Schwarzschild black hole in terms of the Planck mass is given by
\begin{eqnarray}
{M^2}\left( n \right)= 2\left( {n + \frac{1}{2}} \right){m_p}^2
\end{eqnarray}

This result is in agreement with Bekenstein's proposal [27] as we can figure out that the mass of the
BH is proportional to $\sqrt{n}$.

\section{Black hole thermodynamics in phase space with a commutative symplectic structure}

In this section we study thermodynamics of evaporating black holes within a new approach by considering a Hamiltonian model of black hole in phase space with a commutative symplectic structure. This Hamiltonian, as mentioned above, is obtained in Ref. [26]. It is important to note that this Hamiltonian is written in terms of variables that give the dynamics of the spherically symmetric spacetime from viewpoint of a distant observer at rest. By using this quantum Hamiltonian we calculate the partition function of the system. Then other thermodynamical properties are obtained by the partition function straightforwardly.

Suppose that ${\cal{Q}}$ is a configuration space manifold of a mechanical system. Then the Hamiltonian system is defined on cotangent bundle $T^*{\cal{Q}}$  which is the phase space of the system under consideration. Cotangent bundle admits symplectic structure and therefore, the phase space is naturally a symplectic manifold. A
symplectic manifold $(M,\Omega)$ is a manifold $M$ with symplectic structure $\Omega$ which is a closed non-degenerate $2$-form on $M$. The standard Hamiltonian systems are special cases of the above definition. More precisely, a standard Hamiltonian system $(M_{0},\Omega_{0})$ is defined on trivial topology $M_{0} = R^{2n}$. In this commutative, standard chart, the symplectic structure takes the canonical form $\Omega_{0}=dq_{0}^{i}\wedge dp_{0i}$. The variables $(q_{0}^{i}, p_{0i})$ are known as the canonical variables.
The phase space we are dealing with has two canonical variables, $a$ and its momentum $p_a$ with a commutative symplectic structure. As it was mentioned above, the symplectic structure takes the form $\Omega=da\wedge dp_{a}$. The standard local form of Poisson brackets can be deduced in Ref. [28] which leads to the canonical Poisson algebra as
\begin{equation}
\{a,a^\prime\}=0, \quad    \{p_a,a^{\prime}\}=\delta_{a\,a^{\prime}} , \quad     \{p_a,p_{a^\prime}\}=0.
\end{equation}
In such a symplectic structure the volume of the phase space can be written as
\begin{equation}
V(\omega^n)=\int\omega^n=\int d^{n}a\, d^n\,p_a\,.
\end{equation}
So the partition function in this commutative phase space with symplectic structure is given by
\begin{eqnarray}
\begin{array}{l}
{\cal{Z}} = \frac{1}{h}\int {\int_\Gamma  {{e^{\frac{{ - H\left( {{p_a},a} \right)}}{T}}}} da\,d{p_a}} \\
\hspace{0.5cm} = \frac{1}{h}\int_0^{2M} {da} \int_0^\infty  {{e^{\frac{{ - 1}}{T}\left( {\frac{{P_a^2}}{{2a}} + \frac{a}{2}} \right)}}d{p_a}}\,.
\end{array}
\end{eqnarray}
Therefore, we find
\begin{eqnarray}
{\cal{Z}} =-\frac{{2\sqrt {M\pi {T^3}} }}{h}{{\rm{e}}^{ - \frac{M}{T}}} + \frac{{\pi {T^2}}}{h}Erf\left[ {\sqrt {\frac{M}{T}} } \right]\,,
\end{eqnarray}
where $Erf(x)$ is the error function. The Helmholtz free energy of the system can be obtained by the partition function as follows
\begin{eqnarray}
\begin{array}{l}
{\cal{F}} =  - T\ln \left( {\cal{Z}} \right)\\
 \hspace{0.5cm}=  - T\ln \left( { - \frac{{2{{\rm{e}}^{ - \frac{M}{T}}}\sqrt {\pi M{T^3}}  - \pi {T^2}Erf\left( {\sqrt {\frac{M}{T}} } \right)}}{h}} \right)\,.
\end{array}
\end{eqnarray}
The entropy and internal energy of the Schwarzschild black hole in a phase space with commutative symplectic structure take the following forms respectively
\begin{eqnarray}
\begin{array}{l}
{\cal{S}} =  - \frac{{\partial {\cal{F}}}}{{\partial T}} = \frac{{\sqrt {\frac{{M{T^3}}}{T}}  + \sqrt M \left( {2M + 3T} \right)}}{{2\sqrt M T - {{\rm{e}}^{M/T}}\sqrt {\pi {T^3}} Erf\left[ {\sqrt {\frac{M}{T}} } \right]}}\\
\hspace{2cm} - \frac{{2{{\rm{e}}^{M/T}}\sqrt {\pi {T^3}} Erf\left[ {\sqrt {\frac{M}{T}} } \right]}}{{2\sqrt M T - {{\rm{e}}^{M/T}}\sqrt {\pi {T^3}} Erf\left[ {\sqrt {\frac{M}{T}} } \right]}}\\
\hspace{2cm} + \ln \left( { - \frac{{2{{\rm{e}}^{ - \frac{M}{T}}}\sqrt {\pi M{T^3}}  - \pi {T^2}Erf\left[ {\sqrt {\frac{M}{T}} } \right]}}{h}} \right)\,,
\end{array}
\end{eqnarray}

\begin{eqnarray}
\begin{array}{l}
{\cal{U}} =  - {T^2}\frac{\partial }{{\partial T}}(\frac{{\cal{F}}}{T}) =
\frac{{\sqrt M T + \sqrt M \left( {2M + 3T} \right) - 2{{\rm{e}}^{M/T}}\sqrt {\pi {T^3}} Erf\left( {\sqrt {\frac{m}{T}} } \right)}}{{2\sqrt M T - {{\rm{e}}^{M/T}}\sqrt {\pi {T^3}} Erf\left( {\sqrt {\frac{m}{T}} } \right)}}\,.
\end{array}
\end{eqnarray}

We note that this entropy is obtained in a phase space with canonical coordinates. This entropy is calculated from the point of view of a distant observer, not a free falling observer into the black hole. So, it is obvious that this entropy should be different from the standard one obtained in literature for free falling observer into the black hole. In fact, this is the entropy of the external region of a Schwarzschild black hole. As one can see in Figure 1, during evaporation of the black hole, as its temperature increases, the entropy increases more and more due to Hawking radiation.

\begin{figure}
 \begin{center}
\includegraphics{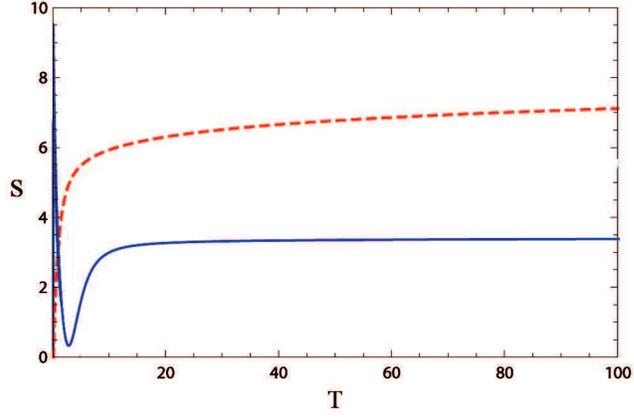}
\end{center}
\vspace{6cm} \caption{\scriptsize{Entropy of black hole versus its temperature. Temperature is in units of
$0.01 T_{p}$ ($T_{p}$ is the Planck temperature) and entropy is in units of Planck energy. The red-dashed curve is calculated in commutative symplectic manifold, while the blue-solid one is for the standard Bekenstein-Hawking case. }}
\end{figure}
We can calculate the heat capacity of the black hole by using the internal energy as follows
\begin{eqnarray}
\begin{array}{l}
{\cal{C}} = \frac{{\partial {\cal{U}}}}{{\partial T}} =  - \frac{{12{M^{5/2}}}}{{{{\left( { - 2M{T^2} + \sqrt {\pi M} {T^{5/2}}{{\rm{e}}^{m/T}}Erf\left[ {\sqrt {\frac{M}{T}} } \right]} \right)}^2}}}\\
 \hspace{1.8cm}- \frac{{{{\rm{e}}^{M/T}}\left( { - 6{M^2}\sqrt {\pi T}  + 4{M^3}\sqrt {\frac{\pi }{T}}  + \left( {M - {M^2}} \right)\sqrt {\pi {T^3}} } \right)Erf\left( {\sqrt {\frac{M}{T}} } \right)}}{{{{\left( { - 2M{T^2} + \sqrt {\pi M} {T^{5/2}}{{\rm{e}}^{m/T}}Erf\left[ {\sqrt {\frac{M}{T}} } \right]} \right)}^2}}}\,.
\end{array}
\end{eqnarray}
Figure 2 shows the behavior of heat capacity versus temperature.
\begin{figure}
  \begin{center}
\includegraphics{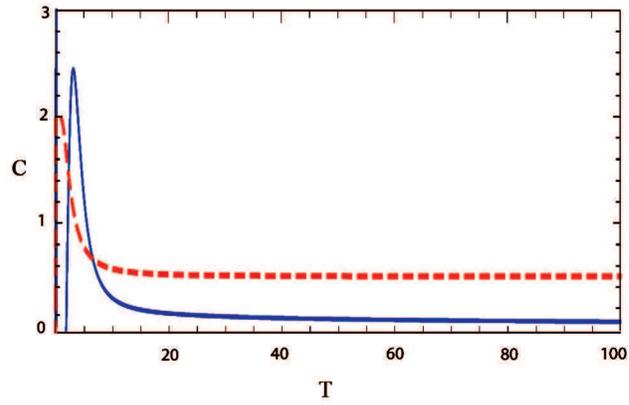}
\end{center}
\vspace{5cm} \caption{\scriptsize{Heat capacity of black hole versus its temperature. The red-dashed curve is calculated in non-commutative symplectic manifold and the blue-solid one is for commutative Snyder phase space.}}
\end{figure}
The results are in accordance with standard approach to black hole thermodynamics, but the only difference is that in this paper we consider a Hamiltonian model for black hole which is based on canonical coordinates from the point of view of a distant observer. So, all the thermodynamical quantities obtained here are for external region of the black hole, far from the event horizon. In this picture a black hole evaporates totally in the final stage of its lifetime. During its evaporation, temperature of the black hole increases and the phase space around the black hole grows up. But the entropy of the black hole decreases due to the Hawking radiation. This happens in the situation where the entropy of the phase space outside the black hole  increases more and more and there is no limit for making it smooth and stopping the evaporation as it's clear from Figure 1.

\section{ Physics of black holes in Snyder noncommutative phase space}
In the pervious section we have considered a non-deformed Hamiltonian system that does not include any cutoff. In this section we consider the Snyder-deformed phase space which induces a UV cutoff for the system under consideration. Inspired by the seminal work of Snyder on discrete spacetime with noncommutative coordinates [21], a locally-deformed noncanonical Poisson algebra is suggested in non-relativistic limit [23,28]. In fact we consider quantum gravity effects as a deformation of the algebra of spacetime. The local form of the symplectic structure in Snyder model is deformed as (see also [20])
\begin{eqnarray}
\omega  = d{q^i} \wedge d{p_i} - \frac{1}{2}d\left( {{q^i}{p_i}} \right) \wedge d\ln \left( {1 + \lambda {p^2}} \right)
\end{eqnarray}
and
\begin{eqnarray}
{\omega ^3} = d{q^1} \wedge d{q^2} \wedge d{q^3} \wedge \frac{{d{p_1} \wedge d{p_2} \wedge d{p_3}}}{{{{\left( {1 + \lambda {p^2}} \right)}^3}}}\,,
\end{eqnarray}
where $\lambda$ is the deformation parameter with dimension of length and is expected to be of the order of the Planck length. Now the partition function for noncommutative case can be interpreted in terms of $a$ and $P_a$ as
\begin{eqnarray}
{{\cal {Z}}_{NC}} &=& \frac{1}{{{h^3}}}\int_\Gamma  {{\omega ^3}{e^{\frac{{ - H\left( {{p_a},a} \right)}}{T}}}}\nonumber\\
&=& \frac{1}{{{h^3}}}\int {{d^3}a\int_\Gamma  {{e^{\frac{{ - H\left( {{p_a},a} \right)}}{T}}}} \frac{{{d^3}{p_a}}}{{{{\left( {1 + \lambda p_a^2} \right)}^3}}}}\,.
\end{eqnarray}
The volume element of the phase space (and therefore integral measure) is changed due to noncommutativity of the Snyder phase space in this case. Note that in another equivalent approach one can deform the variables $a$ and $p_a$ according to the Darboux transformation [20]. The partition function of the Schwarzschild black hole in noncommutative Snyder phase space is given by
\begin{eqnarray}
\begin{array}{l}
{{\cal {Z}}_{NC}}\left( {T,\lambda } \right) = \frac{{\sqrt {{T^3}} {{\rm{e}}^{ - \frac{R}{{2T}}}}}}{h}\left[ {\sqrt {2\pi R} \left( {T{\lambda ^2}\left( {R + 3T} \right) - 1} \right)} \right]\\
\hspace{3.3cm} + \frac{{\pi {T^2}}}{h}\left( {1 - 3{\lambda ^2}{T^2}} \right)Erf\left( {\sqrt {\frac{R}{{2T}}} } \right)\,.
\end{array}
\end{eqnarray}
The entropy, internal energy and heat capacity of the Schwarzschild black hole in Snyder noncommutative phase space can be obtained respectively as follows
\begin{eqnarray}
{\cal {S}}\left( {T,\lambda } \right)= \frac{T}{{{\cal {Z}}_{NC}\left( {T,\lambda } \right)}}\frac{{\partial {\cal {Z}}_{NC}\left( {T,\lambda } \right)}}{{\partial T}} + \ln \left( {{\cal {Z}}_{NC}\left( {T,\lambda } \right)} \right)
\end{eqnarray}

\begin{eqnarray}
{\cal {U}}\left( {T,\lambda } \right) =  - {T^2}\frac{\partial }{{\partial T}}\frac{{{\cal {F}}\left( {T,\lambda } \right)}}{T} = \frac{{{T^2}}}{{{\cal {Z}}_{NC}\left( {T,\lambda } \right)}}\frac{{\partial {\cal {Z}}_{NC}\left( {T,\lambda } \right)}}{{\partial T}}
\end{eqnarray}

\begin{eqnarray}
{\cal {C}} &=& \frac{{2T}}{{{\cal {Z}}_{NC}\left( {T,\lambda } \right)}}\frac{{\partial {\cal {Z}}_{NC}\left( {T,\lambda } \right)}}{{\partial T}}
- {\left( {\frac{T}{{{\cal {Z}}_{NC}\left( {T,\lambda } \right)}}\frac{{\partial {\cal {Z}}_{NC}\left( {T,\lambda } \right)}}{{\partial T}}} \right)^2}\nonumber\\
&+& \frac{{{T^2}}}{{{\cal {Z}}_{NC}\left( {T,\lambda } \right)}}\frac{{{\partial ^2}{\cal {Z}}_{NC}\left( {T,\lambda } \right)}}{{\partial {T^2}}}\,.
\end{eqnarray}

We prefer not to bring explicitly their relations due to lengthy expressions. To compare the results in commutative and noncommutative phase spaces, we have plotted the  modified entropy and heat capacity versus the black hole temperature $T$. In this numerical study, we figured out that by considering the quantum gravity effects encoded as natural UV cutoff in Snyder noncommutative space, at the final stage of evaporation when temperature is in order of the Planck temperature, entropy reaches a limiting value. In fact, when the black hole temperature is so high, all the quantum gravity proposals predict a minimum mass (remnant) of the black hole in which evaporation is stopped. Here, by considering a noncommutative Snyder phase space, we have shown that at high temperatures when the entropy from the point of view of a distant observer is reached to a constant limiting value, the black hole ceases to radiate and a remnant is remained which its entropy is in its minimal value or zero. By considering Fig. 2, we can figure out that the modified heat capacity in Snyder phase space by considering quantum gravity effects is reaching to zero which is in accordance with other approaches to black hole thermodynamics with UV cutoff (see for instance [29,30].

We plotted the deformed entropy in Fig. 3 for some different values of $\lambda$. For large values of $\lambda$ the entropy in the final stage of the black hole evaporation which a distant observer can measure, is decreasing. For $\lambda=0$ , the non-deformed entropy will be recovered.

\begin{figure}
  \begin{center}
\includegraphics{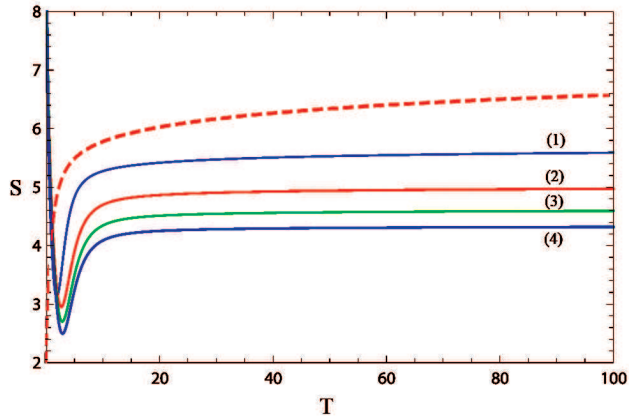}
\end{center}
\vspace{6cm} \caption{\scriptsize{Entropy of black hole versus its temperature for different values of $\lambda$. Temperature is in units of
$0.01 T_{p}$ and entropy is in units of Planck energy. For $\lambda=0$ the entropy for noncommutative symplectic case is shown by red-dashed curve. 
The curves (1), (2), (3) and (4) are for $\lambda=0.1$, $\lambda=0.2$, $\lambda=0.3$ and $\lambda=0.4$, respectively. Increment of $\lambda$ results in reduction of entropy.}}
\end{figure}

\section{ Summery and Conclusion}

In this paper we have studied black hole thermodynamics in a new fashion by considering a symplectic manifold. We have formulated black hole thermodynamics in phase space manifold by using a quantum Hamiltonian and via calculation of the partition function in this space. We note that while usually black hole thermodynamical quantities are calculated from a free falling observer toward black hole horizon, here these thermodynamical properties are calculated in the view of a distant observer at rest. Then we have considered noncommutativity of symplectic manifold by using Snyder noncommutative phase space, so the Poisson algebra on the phase space has changed due to noncommutativity of the phase space. We have shown that a black hole evaporates totally in the final stage of its lifetime in commutative phase space. During its evaporation in commutative case, temperature of the black hole and the volume of the phase space around it goes up but the entropy of the black hole decreases due to the Hawking radiation. In the same time, entropy related to the phase space corresponding to the outside of the black hole increases more and more and there is no limit for making it smooth and stopping the evaporation. In the noncommutative phase space, at high temperatures when the entropy from the point of view of distant observer reaches to a constant limit, the black hole ceases to radiate and a remnant is remained which its entropy is in its minimal value. Finally we note that such a construction is possible in modified theories of gravity such as $f(R)$ theories [31,32,33]. We are going to study this issue in our forthcoming paper. \\

{\bf Acknowledgement}\\
We would like to thank the referee for insightful comments that have boosted the quality of the paper considerably.


\begin{thebibliography}{30}

\bibitem{1}
D. J. Gross and P. F. Mende, Nucl. Phys. B \textbf{303} (1988) 407\\
D. Amati, M. Ciafaloni and G. Veneziano, Phys. Lett. B \textbf{216}  (1989) 41\\
G. Veneziano, Europhys. Lett. \textbf{2} (1986) 199\\
M. Maggiore, Phys. Lett. B {\bf 319} (1993) 83\\
L. Garay, Int. J. Mod. Phys. A \textbf{10},  (1995) 145.

\bibitem{2}
C. Rovelli and L. Smolin, Nucl. Phys. B  \textbf{442}  (1995) 593\\
A. Ashtekar and J. Lewandowski, Class. Quantum Grav. \textbf{14} (1997) A55.

\bibitem{3}
A. Kempf, J. Math. Phys. \textbf{35}  (1994) 4483\\
H. Hinrichsen and A. Kempf, J. Math. Phys. \textbf{37} (1996) 2121\\
A. Kempf, J. Math. Phys. \textbf{38},  (1997) 1347\\
A. Kempf, Phys. Rev. D \textbf{54} (1996) 5174\\
M. Kober and P. Nicolini, Class. Quantum Grav. \textbf{27} (2010)  245024.

\bibitem{4}
V. G. Kadishevsky, Nucl. Phys. B \textbf{141} (1978) 477\\
V. G. Kadishevsky and D. V. Fursaev, Teor. Mat. Fiz. \textbf{83} (1990) 474.

\bibitem{5}
S. Doplicher, K. Fredenhagen and J. E. Roberts, Commun. Math. Phys. \textbf{172} (1995) 187\\
H. Grosse, C. Klimcik and P. Presnajder, Commun. Math. Phys. \textbf{178} (1996) 507\\
H. Grosse, C. Klimcik and P. Presnajder, Int. J. Theor. Phys. \textbf{35} (1996) 231\\
A. Matusis, L. Susskind and N. Toumbas, JHEP  \textbf{0012} (2000) 002\\
I. Chepelev and R. Roiban, JHEP \textbf{0005} (2000) 0375\\
M. Chaichian, A. Demichev and P. Presnajder, Nucl. Phys. B \textbf{567} (2000) 360\\
R. J. Szabo, Phys. Rept. \textbf{378} (2003) 207\\
J. M. Carmona, J. L. Cort\'{e}s, J. Gamboa and F. M\'{e}ndez, JHEP  \textbf{0303} (2003) 058.

\bibitem{6}
A. P. Balachandran, A. Pinzul and B. A. Qureshi, Phys. Lett. B  \textbf{634} (2006) 434.

\bibitem{7}
A. Kempf, G. Mangano and R. B. Mann, Phys. Rev. D  \textbf{52},  (1995) 1108\\
K. Nozari and A. Etemadi, Phys. Rev. D \textbf{85} (2012) 104029.

\bibitem{8}
A. Kempf and G. Mangano, Phys. Rev. D \textbf{55} (1997) 7909\\
A. Smailagic and E. Spallucci, J. Phys. A \textbf{36} (2003) L467\\
A. Smailagic and E. Spallucci, J. Phys. A \textbf{36} (2003) L517\\
R. Garattini and P. Nicolini, Phys. Rev. D \textbf{83} (2011) 064021.

\bibitem{9}
W. Xue, K. Dasgupta and R. Brandenberger, Phys. Rev. D \textbf{83} (2011) 083520.

\bibitem{10}
C. Bastos, O. Bertolami, N. C. Dias and J. N. Prata, J. Math. Phys. \textbf{49} (2008) 072101\\
C. Bastos, N. C. Dias and J. N. Prata, Commun. Math. Phys. \textbf{299} (2010) 709.

\bibitem{11}
N. Seiberg and E. Witten, JHEP \textbf{9909} (1999) 032\\
S. Minwalla, M. V. Raamsdonk and N. Seiberg, JHEP \textbf{0002} (2000) 020.

\bibitem{12}
N. Seiberg, L. Susskind, and N. Toumbas, JHEP \textbf{0006} (2000) 044\\
S. M. Carroll, J. A. Harvey, V. A. Kostelecky, C. D. Lane and T. Okamoto, Phys. Rev. Lett. \textbf{87} (2001)  141601\\
J. M. Carmona, J. L. Cort\'{e}s, J. Gamboa and F. M\'{e}ndez, Phys. Lett. B \textbf{565} (2003) 222.

\bibitem{13}
A. P. Balachandran, T. R. Govindarajan, G. Mangano, A. Pinzul, B. A. Qureshi and S. Vaidya, Phys. Rev. D \textbf{75} (2007) 045009\\
A. Pinzul, J. Phys. A \textbf{45} (2012) 075401.

\bibitem{14}
K. Konishi, G. Paffuti and P. Provero, Phys. Lett. B \textbf{234} (1990) 276\\
M. Maggiore, Phys. Rev. D \textbf{49} (1994) 5182\\
S. Hossenfelder, Living Rev. Relativity {\bf 16} (2013) 2.

\bibitem{15}
A. Ashtekar, S. Fairhurst and J. Willis, Class. Quantum Grav. \textbf{20} (2003) 1031\\
K. Fredenhagen and F. Reszewski, Class. Quantum Grav. \textbf{23} (2006) 6577.

\bibitem{16}
A. Corichi, T. Vukasinac and J. A. Zapata, Phys. Rev. D \textbf{76} (2007) 044016.

\bibitem{17}
M. V. Battisti and S. Meljanac, Phys. Rev. D  \textbf{79} (2009) 067505\\
M. V. Battisti, Phys. Rev. D \textbf{79} (2009) 083506\\
S. Das, M. P. G. Robbins and M. A. Walton, Can. J. Phys. \textbf{94} (2016) 139, [arXiv:1412.6467 [gr-qc]].

\bibitem{18}
M. A. Gorji, K. Nozari and B. Vakili, Class. Quantum Grav. \textbf{32} (2015) 155007.

\bibitem{19}
M. A. Gorji, K. Nozari and B. Vakili, Phys. Rev. D \textbf{89} (2014) 084072.

\bibitem{20}
K. Nozari, M. A. Gorji, V. Hosseinzadeh and B. Vakili, Class. Quantum. Grav. \textbf{33} (2016) 025009.

\bibitem{21}
H. Snyder, Phys. Rev. \textbf{71} (1947) 38.

\bibitem{22}
S. Mignemi, Phys. Rev. D  \textbf{84} (2011)  025021.

\bibitem{23}
S. Mignemi, [arXiv:1110.0201 [hep-th]].

\bibitem{24}
B. Ivetic, S. Melijanac and S. Mignemi, Class. Quantum Grav. \textbf{31} (2014) 105010.

\bibitem{25}
S. Mignemi, R. Strajn, Phys. Rev. D \textbf{90} (2014) 044019.

\bibitem{26}
J. M\"{a}kel\"{a} and P. Repo, Phys. Rev. D \textbf{57} (1998) 4899.

\bibitem{27}
J. D. Bekenstein, Lett. Nuovo Cimento \textbf{11} (1974) 467.

\bibitem{28}
S. Mignemi, Class. Quantum Grav. \textbf{29} (2012) 215019.

\bibitem{29}
K. Nozari and S. Saghafi, JHEP \textbf{11} (2012) 005\\
W. Kim, E. J. Son and M. Yoon, JHEP \textbf{0801} (2008) 035.

\bibitem{30}
K. Nozari and S. H. Mehdipour, Class. Quant. Grav. \textbf{25} (2008) 175015\\
K. Nozari, S. Saghafi and A. D. Kamali, Astrophys Space Sci, \textbf{357} (2015) 140\\
H. Soltani, A. D. Kamali and K. Nozari, Advances in High Energy Physics,
Volume 2014, Article ID 247208, http://dx.doi.org/10.1155/2014/247208.

\bibitem{31}
S. Nojiri and S. D. Odintsov, Phys. Rept. \textbf{505} (2011) 59.

\bibitem{32}
S. Capozziello and M. De Laurentis, Phys. Rept. \textbf{509}, (2011) 167.

\bibitem{33}
S. Nojiri, S. D. Odintsov and V. K. Oikonomou, Phys. Rept. in press, [arXiv:1705.11098].


\end{thebibliography}
\end{document}